\documentclass[epj]{svjour}

\usepackage{lineno}
\usepackage{multirow}
\usepackage{amssymb}
\usepackage{amsmath}
\usepackage{graphicx,color}
\usepackage{url}
\usepackage[colorlinks,linkcolor=blue,citecolor=blue,filecolor=black,urlcolor=blue]{hyperref}
\usepackage{harpoon}
\usepackage{url}
\usepackage{xspace}	
\usepackage[T1]{fontenc}
\usepackage{booktabs}
\usepackage{appendix}
\usepackage{orcidlink}
\usepackage[dvipsnames]{xcolor}
\usepackage{subcaption}
\usepackage[labelformat=parens, labelfont=bf]{subcaption}
\usepackage{microtype}
\usepackage{cite}
\begin{document}

\title{Flavour-Dependent Chemical Freeze-Out of Light Nuclei in Relativistic Heavy-Ion Collisions}

\author{%
  Rishabh Sharma\protect\orcidlink{0000-0003-0698-7363}\inst{1}
  \and
  Fernando Antonio Flor\protect\orcidlink{0000-0002-0194-1318}\inst{2}\thanks{\emph{Current address:} Physics Division, Argonne National Laboratory, Lemont, IL, USA 60439}
  \and
  Sibaram Behera\protect\orcidlink{0009-0004-7441-0472}\inst{1}
  \and
  Chitrasen Jena\protect\orcidlink{0000-0003-2270-3324}\inst{1}\thanks{\emph{Email address:} \href{mailto:cjena@iisertirupati.ac.in}{cjena@iisertirupati.ac.in}}
  \and
  Helen Caines\protect\orcidlink{0000-0002-1595-411X}\inst{2}
}                    
\offprints{}          
\institute{Department of Physics, Indian Institute of Science Education and
Research (IISER) Tirupati, Tirupati, 517619, Andhra Pradesh, India \and Wright Laboratory, Yale University, New Haven, CT, 06520, USA}
%
%
\abstract{
We study the production of light nuclei in Au+Au collisions at
$\sqrt{s_\mathrm{NN}}$ = 7.7 -- 200 GeV and Pb+Pb collisions at $\sqrt{s_\mathrm{NN}}$ = 2.76 and 5.02 TeV within a flavour-dependent chemical freeze-out scenario, assuming different flavoured hadrons undergo separate chemical freeze-out. Using the \texttt{Thermal-FIST} package, thermal parameters extracted from fits to various sets of hadron yields, including and excluding light nuclei, are used to calculate the ratios of the yields of light nuclei, namely, $d/p$, $\bar{d}/\bar{p}$, $t/p$, $t/d$, $^4\text{He}/^3\text{He}$, and $^3\text{H}_{\Lambda}/^3\text{He}$. A comparison with experimental data from the STAR and ALICE collaborations shows that a sequential freeze-out scenario provides a better description of light nuclei yield ratios than the traditional single freeze-out approach. These results suggest the flavour-dependent chemical freeze-out for final state light nuclei production persists in heavy-ion collisions at both RHIC and LHC energies.}
%
%
\authorrunning{\textit{R. Sharma}, et al.}
\titlerunning{\textit{Flavour-Dependent Chemical Freeze-Out of Light Nuclei in Relativistic Heavy-Ion Collisions}}
\maketitle

\section{Introduction}

\label{sec:intro}
In relativistic heavy-ion collisions, nuclear matter is heated to extreme temperatures and energy densities, forming a deconfined state of quarks and gluons known as the quark-gluon plasma (QGP) \cite{STAR:2005gfr,Aoki:2006we,Ejiri:2008xt,ALICE:2022wpn}. The high energy density created in the collisions results in a strong pressure gradient, causing a rapid expansion and cooling of the produced QGP. As the system evolves, it transitions to an interacting gas of hadrons. This hadronic "fireball" continues to expand and eventually undergoes chemical freeze-out, where inelastic collisions cease, fixing the relative yields of stable hadrons. Elastic collisions after chemical freeze-out keep the system in thermal equilibrium until kinetic freeze-out \cite{Chatterjee:2015fua}.

The Hadron Resonance Gas (HRG) model has been widely used to study hadron production in heavy-ion collisions \cite{Cleymans:1998fq,Becattini:2009sc,Andronic:2017pug}. In its simplest formulation, with only a few thermal parameters—volume (or radius), chemical freeze-out temperature ($T_\text{ch}$), and chemical potentials—it has been remarkably successful in describing the yields of various hadrons produced across a wide range of collision systems and energies. A comparison of transverse momentum ($p_T$)-integrated hadron yields ($dN/dy$) with the predictions from the HRG allows the estimation of these parameters and defines the thermodynamic state of the system at chemical freeze-out.

Traditionally, chemical freeze-out is understood as the process where all the hadrons in the fireball freeze-out at the same temperature (1CFO). However, this process can be more complex where particles may freeze out at different temperatures. Since temperature decreases over time in the expanding fireball, different freeze-out temperatures imply a sequential freeze-out. Hadrochemical analyses \cite{Chatterjee:2013yga,Bugaev:2013sfa,Chatterjee:2014ysa} and high-precision continuum-extrapolated lattice QCD calculations \cite{Ratti:2011au,Bellwied:2013cta} suggest a flavour-dependent chemical freeze-out scenario in the QCD phase transition crossover region. This implies that quark flavours, such as light and strange quarks, may freeze out at different temperatures, establishing a flavour hierarchy during the transition. Scenarios where strange-flavour hadrons freeze out earlier than light-flavour hadrons (2CFO) have been studied and found to describe the experimental data better compared to a 1CFO scenario \cite{Chatterjee:2013yga,Flor:2020fdw,Flor:2021olm}. It has been reported that $T_\text{ch}$ for light hadrons is approximately 150.2 $\pm$ 2.6 MeV, while for strange hadrons it is around 165.1 $\pm$ 2.7 MeV, at vanishing baryon chemical potential \cite{Flor:2020fdw}.

Although the HRG model has shown good agreement with the $dN/dy$ of light nuclei such as $d$, $t$, $^3\text{He}$, and $^4\text{He}$ at LHC energies, the underlying production mechanism remains unclear \cite{Andronic:2010qu,Yu:2024sof}. Due to their low binding energies (of the order of a few MeV), it remains puzzling how such composite objects could survive in a hadronic fireball, where the chemical freeze-out temperatures (150-160 MeV) are well above these energy scales \cite{Oliinychenko:2020ply}. However, it has been argued that the relative yield of nuclei is determined by the entropy per baryon, which is fixed at chemical freeze-out \cite{Andronic:2010qu,Siemens:1979dz,Hahn:1986mb}. Therefore, entropy conservation governs the production yields of light nuclei. Recently, at RHIC energies, the STAR collaboration reported that although the thermal model describes the $d/p$ yield ratio, it systematically overestimates the $t/p$ ratio \cite{STAR:2019sjh,STAR:2022hbp}. Notably, hadronic rescattering has been proposed as a key mechanism for explaining light nuclei yields, particularly those of tritons, at RHIC and LHC energies. When these effects are incorporated, the observed triton yields become consistent with thermal model predictions~\cite{Sun:2022xjr}.

A recent study of light nuclei production at RHIC energies indicated a distinct freeze-out of light nuclei, reporting freeze-out temperatures of 150.2 $\pm$ 6 MeV for light hadrons, 165.1 $\pm$ 2.7 MeV for strange hadrons, and 141.7 $\pm$ 1.4 MeV for light nuclei \cite{Yu:2024sof}. Given the similarity between the freeze-out temperature of light nuclei and that of light hadrons, it is reasonable to assert that light nuclei form near the chemical freeze-out of light hadrons in the 2CFO scenario, as shown in \cite{Chatterjee:2014ysa}.

In this letter, we study the impact of the 2CFO scenario on the production of light nuclei. Chemical freeze-out parameters were estimated by performing thermal fits to the $dN/dy$ of various hadrons in the most central Au+Au collisions at $\sqrt{s_\mathrm{NN}}$ = 7.7 -- 200 GeV and Pb+Pb collisions at $\sqrt{s_\mathrm{NN}}$ = 2.76 and 5.02 TeV. We explore how the chemical freeze-out parameters are affected by the inclusion of light nuclei, $d(\overline{d})$, $t$, $^3\text{He}$($^3\overline{\text{He}}$), in the thermal fits. In addition, light nuclei ratios calculated in both 1CFO and 2CFO scenarios were compared with experimental data from the STAR and ALICE collaborations. Furthermore, we investigate the effects of considering partial chemical equilibrium (PCE) in the HRG model framework. In the context of heavy-ion collisions, PCE is defined as a state where short-lived resonances are allowed to decay and regenerate while maintaining a quasi-equilibrium with the stable hadrons produced at chemical freeze-out~\cite{Motornenko:2019jha}. This approach provides a more refined description of post-chemical freeze-out expansion of hadronic medium in heavy-ion collisions. Light nuclei can be incorporated into the HRG-PCE framework using the Saha equation, drawing an analogy from big bang nucleosynthesis~\cite{Vovchenko:2019aoz}. Compared to previous studies (e.g., Refs.~\cite{Flor:2020fdw,Flor:2021olm}), a key advancement of this present work is the simultaneous inclusion of $d$ and $t$ (or $^3\text{He}$) with rest of the hadrons in the thermal fits, along with the incorporation of PCE, an aspect not considered in earlier analyses limited to a simple HRG model. In addition, this study employs the energy-dependent Breit-Wigner (eBW) scheme for treating resonance decays, instead of the zero-width approximation, leading to a more realistic description of resonance contributions to final hadron yields~\cite{Vovchenko:2018fmh}. The letter is organized as follows: Sec. \ref{sec:model} describes the HRG model, Sec. \ref{sec:results} discusses our findings with respect to light nuclei production, and we present a summary of our results in Sec. \ref{sec:summary}.

\begin{figure*}
    \centering

    \begin{subfigure}[b]{1.0\textwidth}
        \centering
        \includegraphics[width=\textwidth]{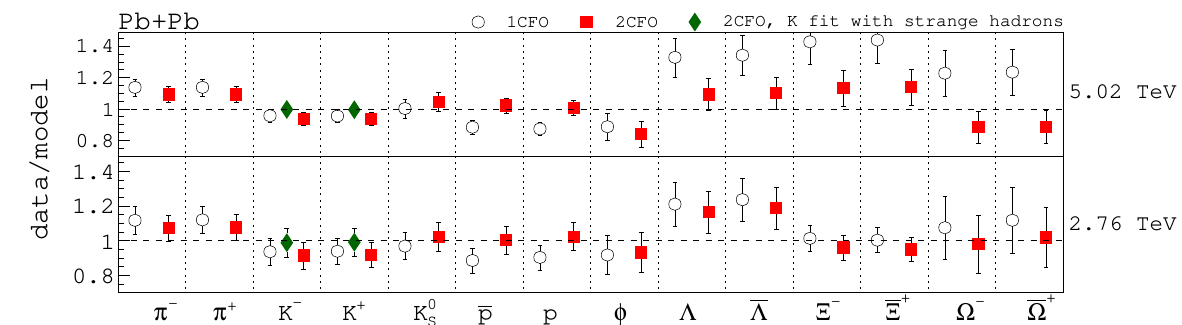}
        \caption*{}
    \end{subfigure}

    \vspace{0.1em}

    \begin{subfigure}[b]{1.0\textwidth}
        \centering
        \includegraphics[width=\textwidth]{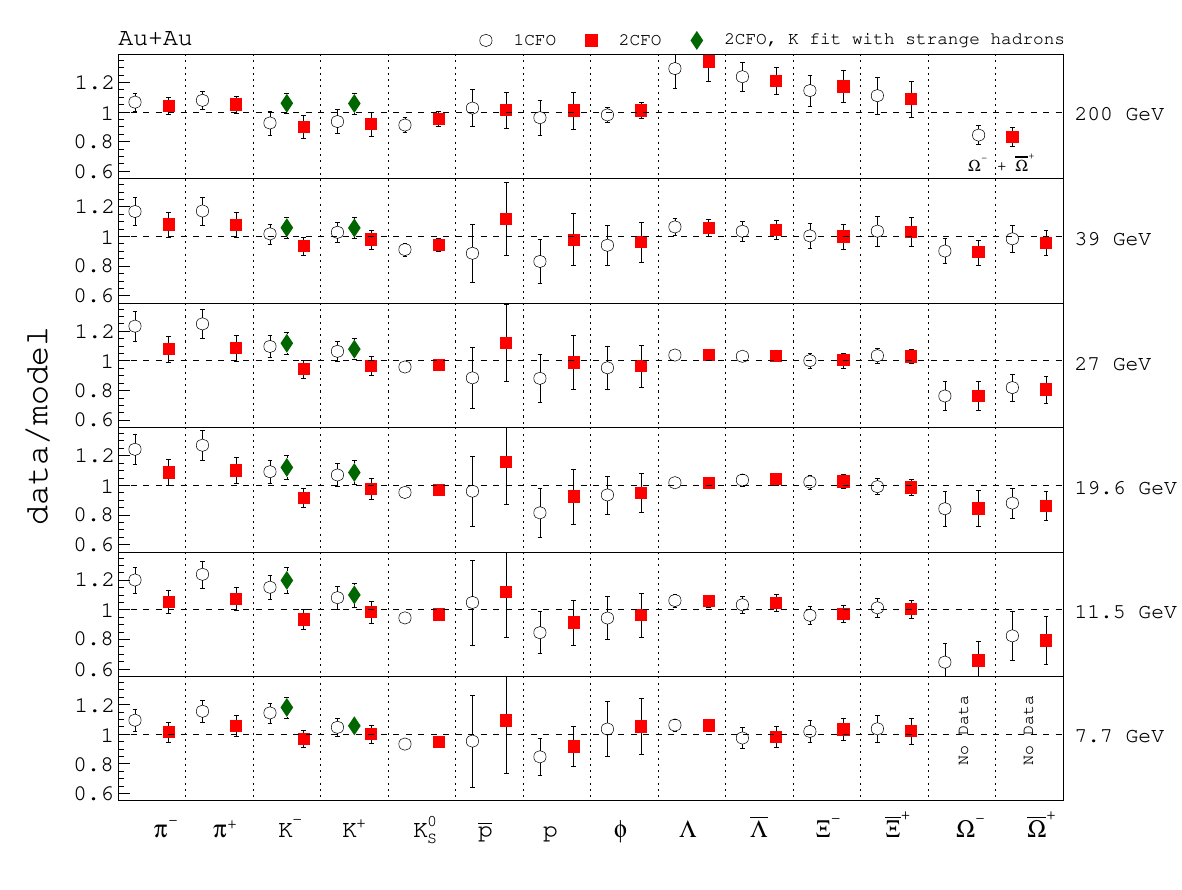}
        \caption*{}
    \end{subfigure}

\caption{Ratios of experimental data to thermal model fit to $\pi$, $K$, $K_s^{0}$, $p$, $\phi$, $\Lambda$, $\Xi$, and $\Omega$ in 0--10\% centrality of Au+Au and Pb+Pb collisions at $\sqrt{s_\mathrm{NN}}$ = 7.7 – 5020 GeV ($K^0_s$, $\Lambda$, $\Xi$, and $\Omega$ yields at $\sqrt{s_\mathrm{NN}}$ = 200 GeV were measured in 0--5\% centrality). The 1CFO yield calculations are shown as open black circles while the 2CFO calculations are shown as solid red squares. The solid green diamonds indicate the case where the charged kaons were included in the \textit{strange hadrons} set.}
    \label{fig:Fit}
\end{figure*}

\begin{figure*}

    \begin{subfigure}[b]{1.0\textwidth}
        \hspace*{-0.09\textwidth}
        \includegraphics[width=1.2\textwidth]{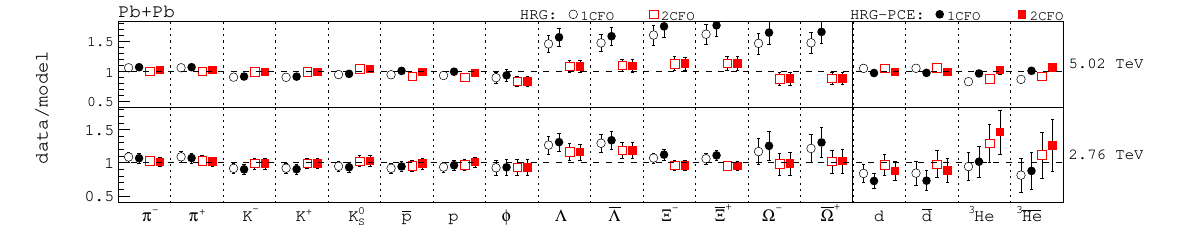}
        \caption*{}
    \end{subfigure}

    \vspace{0.1em}

    \begin{subfigure}[b]{1.0\textwidth}
        \hspace*{-0.09\textwidth}
        \includegraphics[width=1.2\textwidth]{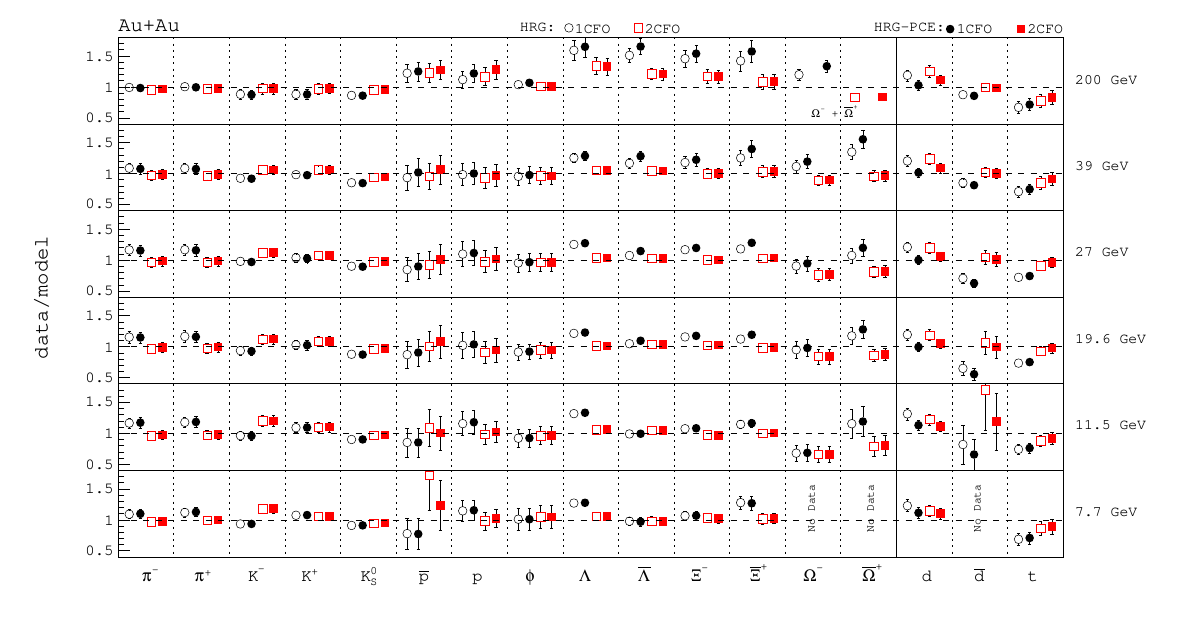}
        \caption*{}
    \end{subfigure}

\caption{Ratios of experimental data to thermal model fit to $\pi$, $K$, $K_s^{0}$, $p$, $\phi$, $\Lambda$, $\Xi$, $\Omega$, $d$, and $t$ (or $^3\text{He}$) in 0--10\% centrality of Au+Au and Pb+Pb collisions at $\sqrt{s_\mathrm{NN}}$ = 7.7 - 5020 GeV ($K^0_s$, $\Lambda$, $\Xi$, and $\Omega$ yields at $\sqrt{s_\mathrm{NN}}$ = 200 GeV were measured in 0--5\% centrality). Thermal model calculations are shown for: 1CFO in HRG (open black circles), 2CFO in HRG (open red squares), 1CFO in HRG-PCE (solid black circles), and 2CFO in HRG-PCE (solid red squares).}
    \label{fig:Fit_nuclei}
\end{figure*}

\begin{figure}[!htbp]
    \centering
    \includegraphics[width=0.50\textwidth]{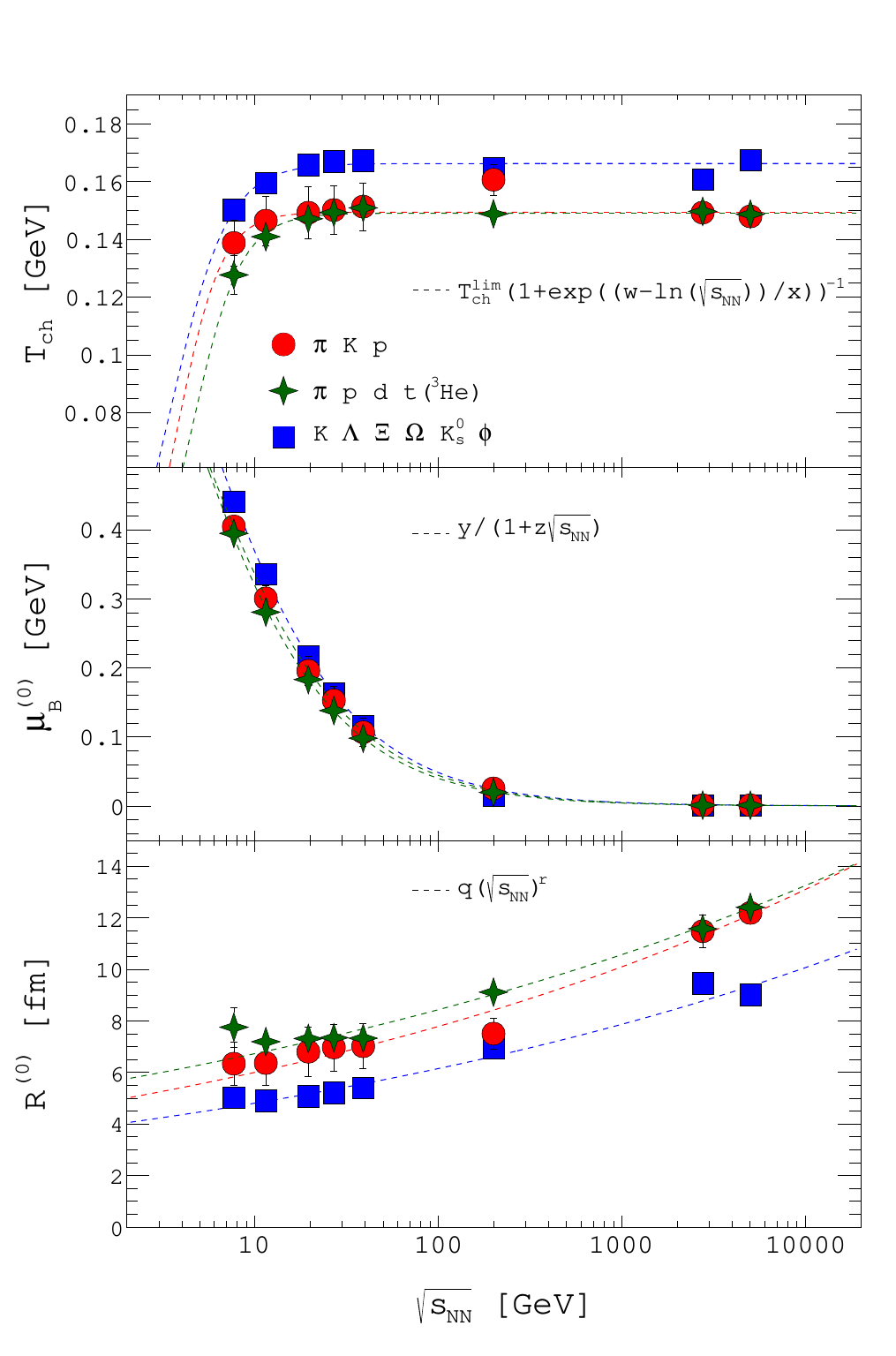}
    \caption{Chemical freeze-out parameters $T_\text{ch}$, $\mu^{(0)}_B$, and $R^{(0)}$ in the 2CFO scenario, extracted from thermal fits to \textit{light hadrons} (red circles), \textit{light hadrons + nuclei} (green double diamonds), and \textit{strange hadrons} (blue squares).}
    \label{fig:para_en}
\end{figure}

\begin{table*}[!htb]
\centering
\caption{Fit parameters from parametrization of $T_{\text{ch}}$, $\mu^{(0)}_B$, and $R^{(0)}$ for different particle sets.}
\begin{tabular}{@{}lccccc@{}}
\toprule
Parameters  & \multicolumn{2}{c}{1CFO}                     & \multicolumn{3}{c}{2CFO}                              \\
                    \cmidrule(lr){2-3} \cmidrule(l){4-6}
            & \textit{hadrons}   & \textit{hadrons}+\textit{nuclei} & \textit{light hadrons}   & \textit{light hadrons}+\textit{nuclei} & \textit{strange hadrons} \\ \cmidrule(lr){2-3} \cmidrule(l){4-6}
{$T_{\text{ch}}^{\text{lim}}$ (MeV)} 
& $162.89 \pm 0.67$ & $152.72 \pm 0.28$ & $149.41 \pm 1.50$          & $149.05 \pm 0.40$        & $166.72 \pm 0.88$         \\
{$w$} 
& 1.47              & 1.55              & 1.33                       & 1.51                     & 1.28                     \\
{$x$} 
& 0.24              & 0.19              & 0.28                       & 0.31                     & 0.35                     \\
{$y$ (GeV)} 
& 1.38              & 1.41               & 1.26                      & 1.45                     & 1.40                     \\
{$z$ (GeV$^{-1}$)} 
& 0.28              & 0.36               & 0.27                      & 0.35                     & 0.28
                 \\
{$q$ (fm)} 
& 3.56              & 4.36               & 4.62                      & 5.36                     & 3.72                     \\
{$r$} 
& 0.13              & 0.12               & 0.11                      & 0.10                     & 0.11
\\ \bottomrule
\end{tabular}
\label{tab:fit_parameters}
\end{table*}

\begin{figure*}

    \begin{subfigure}[b]{0.97\textwidth}
        \includegraphics[width=1.0\textwidth]{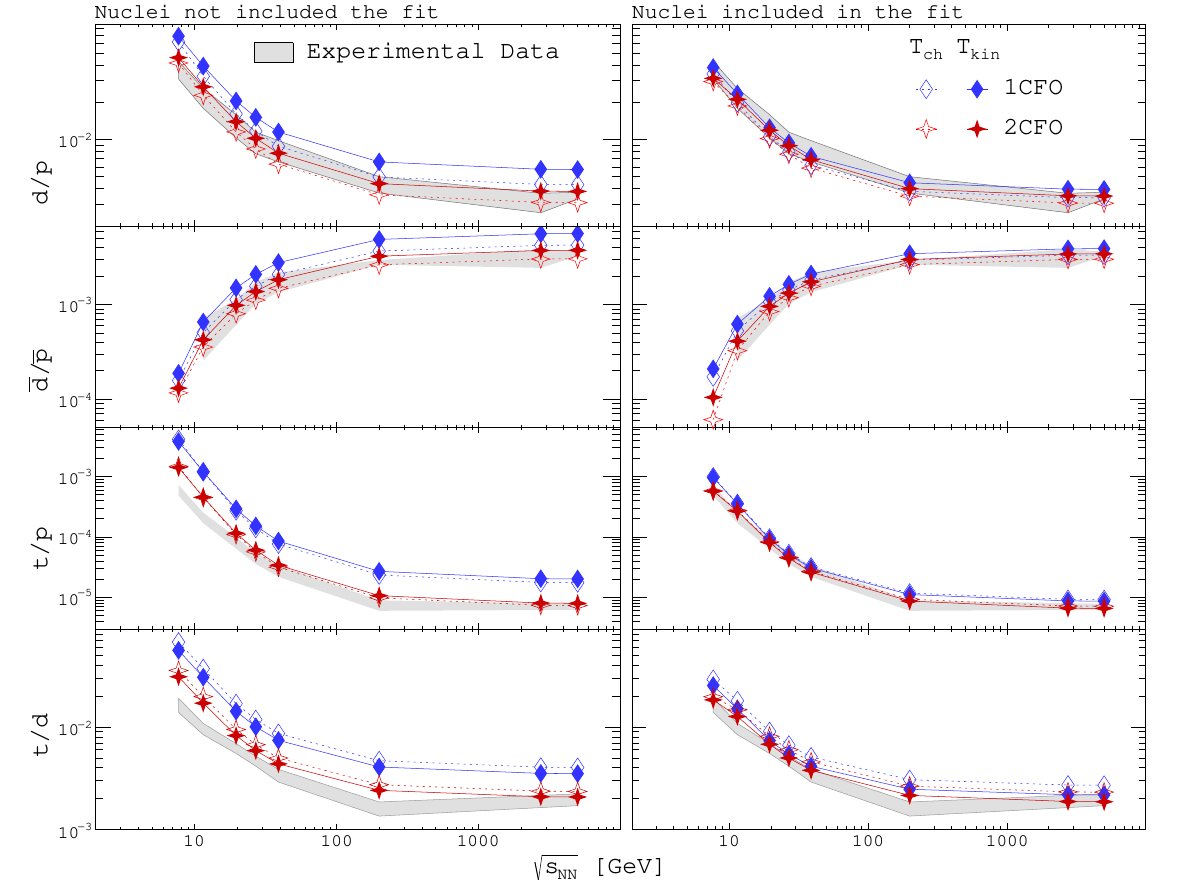}
        \caption*{}
    \end{subfigure}


    \begin{subfigure}[b]{0.97\textwidth}
        \includegraphics[width=1.0\textwidth]{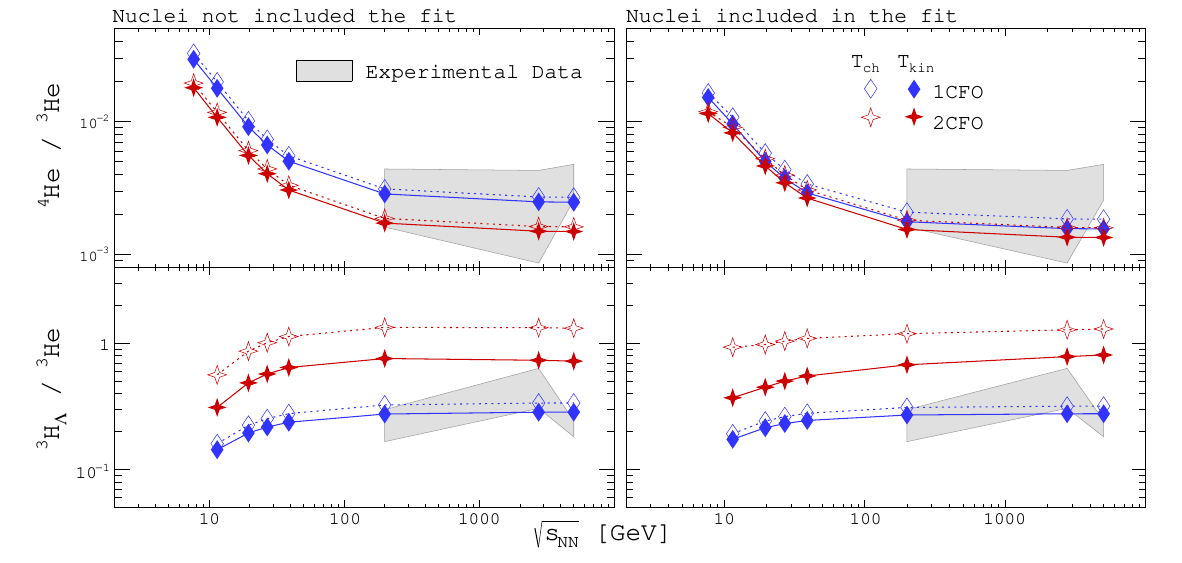}
        \caption*{}
    \end{subfigure}

\caption{Comparison of light nuclei yield ratios—$d/p$, $\bar{d}/\bar{p}$, $t/p$, $t/d$, $^4\text{He}/^3\text{He}$, and $^3\text{H}_{\Lambda}/^3\text{He}$—with thermal model calculations at $T = T_\text{ch}$ and $T = T_{kin} = 100$ MeV, open and full markers, respectively. The calculations use chemical freeze-out parameters from Eq.~\ref{eq:part} obtained under both the 1CFO and 2CFO scenarios, blue and red markers, respectively, without (left) and with (right) the inclusion of light nuclei in the thermal fits.}
    \label{fig:LNRatios}
\end{figure*}
\section{Hadron Resonance Gas Model}
\label{sec:model}
The analysis in this letter was performed using the open-source \texttt{Thermal-FIST} (Thermal, Fast and Interactive Statistical Toolkit) package, a versatile framework designed for studying particle production in high-energy collisions within the HRG model \cite{Vovchenko:2019pjl}. The \texttt{Thermal-FIST} offers a range of capabilities, including the computation of hadron yields, ratios, and fluctuations and the ability to perform thermal fits using the Canonical Ensemble and the Grand Canonical Ensemble (GCE). Additionally, the \texttt{Thermal-FIST} supports calculations within the PCE framework, allowing for the modeling of post-chemical freeze-out evolution of hadrnoic medium~\cite{Motornenko:2019jha}. In our analysis, we used the GCE framework, where the conserved QCD charges—baryon number ($B$), electric charge ($Q$), and strangeness ($S$)—are not conserved strictly but are maintained on average through their corresponding chemical potentials, $\mu^{(0)}_B$, $\mu^{(0)}_Q$, and $\mu^{(0)}_S$, respectively. The primordial yield of the $i^{\text{th}}$ particle in the GCE at chemical freeze-out is given by:

\begin{equation} 
N_{i,\text{prim}}^{(0)} = \frac{g_i V^{(0)}}{2\pi^2} \int_0^\infty \frac{p^2 dp}{\exp{\left(\frac{E_i - \mu_i^{(0)}}{T_{\text{ch}}}\right)} \pm 1}, 
\end{equation}

where $g_i$ is the degeneracy factor, $V^{(0)}=\frac43\pi (R^{(0)})^3$ is the system volume at chemical freeze-out with radius $R^{(0)}$, $p$ is the particle momentum, $E_i = \sqrt{p^2 + m_i^2}$ is the energy, and $\mu^{(0)}_i = B\mu^{(0)}_B + Q\mu^{(0)}_Q + S\mu^{(0)}_S$ is the chemical potential. The $+$ and $-$ signs correspond to fermions and bosons, respectively. The superscript $(0)$ is used to denote quantities associated with chemical freeze-out.

The final yield of the $i^{\text{th}}$ particle species is calculated as~\cite{Vovchenko:2018fmh}:
\begin{equation}
    N^{(0)}_{i,\text{tot}} = N^{(0)}_{i,\text{prim}} + \sum_{\mathcal{R}} n_{i,\mathcal{R}}N_{\mathcal{R},\text{prim}}
    \label{eq:HRG-Yield}
\end{equation}
where $n_{i,\mathcal{R}}$ is the average number of particles of type $i$ resulting from the decay of the resonance $\mathcal{R}$, and $N_{\mathcal{R},\text{prim}}$ is the primordial yield of $\mathcal{R}$. In the PCE framework, light nuclei can continue to evolve chemically beyond $T_{\text{ch}}$, undergoing reactions that maintain a relative chemical equilibrium with their constituent particles. Due to their large break-up cross-sections, light nuclei $A$ can be involved in reactions of the form $X + A \leftrightarrow X + \sum_i A_i$, where $A_i$ are the components of the nucleus (such as protons, neutrons, hyperons, or lighter nuclei), and $X$ represents some hadron (e.g., a pion)~\cite{Garcilazo:1987rx}. The evolution of nuclei yields in this framework follows the Saha equation, ensuring that their densities remain in equilibrium with those of their constituents~\cite{Vovchenko:2019aoz}:

\begin{equation}
\frac{n_{A,\text{tot}}}{\prod_i n_{{A_i,\text{tot}}}} = \frac{n_{A,\text{tot}}^{(0)}}{\prod_i n_{{A_i,\text{tot}}}^{(0)}},
\end{equation}

where $n_{A,tot}$ and $n_{A_i,tot}$ denote the number densities of the nucleus and its components, respectively. This relation implies $\mu_A = \sum_{i} \mu_{A_i}$. At temperatures $T < T_{\text{ch}}$, the yield of a nucleus $A$ is modified as:

\begin{equation}
N_{A,\text{tot}}(T) = N_{A,\text{tot}}^{(0)} \exp\left(\frac{\mu_A}{T}\right) \frac{V}{V^{(0)}},
\label{eq:HRG-PCE-Yield}
\end{equation}

where $\mu_A$ and the system volume $V$ are determined at a temperature $T<T_\text{ch}$ under the assumption of isentropic expansion, ensuring the conservation of the total stable hadron yields within the HRG-PCE framework.

We used the PDG2020 \cite{ParticleDataGroup:2020ssz} hadronic spectrum available in \texttt{Thermal-FIST} v1.4.2. The $dN/dy$ of identified hadrons ($\pi^{\pm}$, $K^{\pm}$, $K^0_s$, $p(\overline{p})$, $\phi$, $\Lambda(\overline{\Lambda})$, $\Xi^{-}(\overline{\Xi}^+)$, and $\Omega^{-}$ ($\overline{\Omega}^+$)) and light nuclei ($d(\overline{d})$, and $t$ or $^3\text{He}$($^3\overline{\text{He}}$)) in 0--10\% central Au+Au collisions at $\sqrt{s_\mathrm{NN}}$ = 7.7, 11.5, 19.6, 27, 39, and 200 GeV and Pb+Pb collisions at $\sqrt{s_\mathrm{NN}}$ = 2.76 and 5.02 TeV were used in the thermal fits from the STAR and ALICE collaborations \cite{STAR:2019sjh,STAR:2022hbp,STAR:2004yym,STAR:2019bjj,STAR:2017sal,Bellini:2018khg,ALICE:2022veq,ALICE:2019hno,ALICE:2014jbq,ALICE:2023ulv,STAR:2008med,STAR:2011fbd}. For Au+Au collisions at $\sqrt{s_\mathrm{NN}}$ = 200 GeV, the 0--5\% values were used for $K^0_s$, $\Lambda(\overline{\Lambda})$, $\Xi^{-}(\overline{\Xi}^+)$, and $\Omega^{-}$ + $\overline{\Omega}^+$ yields due to the binning of the experimental data. For brevity, we adopt a shorthand notation where the symbol of a particle (e.g. $p$) represents both the particle ($p$) and its corresponding anti-particle ($\overline{p}$). This convention is followed throughout this letter unless explicitly stated otherwise. 

The free parameters in our thermal fits are $ T_{\text{ch}} $, $ \mu_B^{(0)} $, and $ R^{(0)} $; however, at LHC energies, $ \mu_B^{(0)} $ is fixed to 1\,MeV due to the approximate baryon--antibaryon symmetry. The chemical potentials $\mu_Q$ and $\mu_S$ were constrained using the conditions, net $Q/B = 0.4$ and net $S = 0$, ensuring conservation of electric charge and strangeness. We set the strangeness fugacity factor, $\gamma_s = 1$, assuming complete strangeness equilibrium and we applied quantum statistics to all particles. Finite resonance widths were incorporated using an energy-dependent Breit-Wigner (eBW) distribution that more realistically considers mass-dependent resonance widths, impacting hadrons yields, particularly for protons with reduced feed-down from broad $\Delta$ resonances~\cite{Vovchenko:2018fmh}. The particle sets considered in the analysis are as follows:
\begin{itemize}
    \item \textbf{1CFO scenario}: 
    \begin{itemize}
        \item $\pi$, $K$, $K_s^{0}$, $p$, $\phi$, $\Lambda$, $\Xi$, and $\Omega$  [\textit{hadrons}]
        \item $\pi$, $K$, $K_s^{0}$, $p$, $\phi$, $\Lambda$, $\Xi$, $\Omega$, $d$, and $t$ or $^3\text{He}$
        
        [\textit{hadrons}+\textit{nuclei}]
    \end{itemize}
    \item \textbf{2CFO scenario}:
    \begin{itemize}
        \item $\pi$, $K$, and $p$ [\textit{light hadrons}]
        \item $\pi$, $p$, $d$, and $t$ 
 or $^3\text{He}$ [\textit{light hadrons}+\textit{nuclei}]
        \item $K$, $K_s^{0}$, $\phi$, $\Lambda$, $\Xi$, and $\Omega$  [\textit{strange hadrons}]
    \end{itemize}
\end{itemize}

Including $K$ in the \textit{light hadrons} fit prevents too few statistical degrees of freedom, as it has been shown that it does not significantly impact $T_\text{ch}$ \cite{Magestro:2001jz,Flor:2020fdw}. Furthermore, to study the effect of light nuclei on the chemical freeze-out, we performed thermal fits to various particle species, incorporating light nuclei as detailed above. When including light nuclei in the fits, we considered two approaches:
\begin{itemize}
    \item \textbf{HRG}: Light nuclei yields, like those of hadron, are computed at $T_{\text{ch}}$ using Eq.~(\ref{eq:HRG-Yield}).
    
    \item \textbf{HRG-PCE}: While the stable hadrons yields are calculated at $T_{\text{ch}}$ using Eq.~(\ref{eq:HRG-Yield}), light nuclei yields are calculated at a lower temperature, $T_{\text{kin}} = 100$ MeV using Eq.~(\ref{eq:HRG-PCE-Yield}). This approach does not imply that light nuclei freeze out at $T_{\text{kin}}$, but rather models their effective yields in a hadronic medium evolving from $T_{\text{ch}}$ and $T_{\text{kin}}$. The choice of $T_{\text{kin}}$ is motivated by blast-wave model~\cite{Schnedermann:1993ws} analyses, which typically suggest $T_{\text{kin}} \approx 100$ MeV for most-central heavy-ion collisions~\cite{STAR:2017sal,ALICE:2015wav,NA49:2016qvu}.
\end{itemize}

This distinction allows us to explore the impact of post-freeze-out hadronic phase evolution on final-state light nuclei yields. The freeze-out parameters obtained from the thermal fits to the \textit{hadrons}(+\textit{nuclei}) and \textit{light hadrons}(+\textit{nuclei}) particle sets were used to calculate the nuclei yield ratios, $d/p$, $\bar{d}/\bar{p}$, $t/p$, $t/d$, $^4\text{He}/^3\text{He}$, and $^3\text{H}_{\Lambda}/^3\text{He}$ . The results from the 1CFO and 2CFO calculations are compared to determine which scenario best describes the experimentally reported nuclei yield ratios.

\section{Results}
\label{sec:results}

Figure~\ref{fig:Fit} shows the ratios of experimental yields to the corresponding thermal model values obtained from fits performed without including light nuclei, using both the 1CFO and 2CFO scenarios within the HRG framework. These fits cover Au+Au collisions at $\sqrt{s_{\mathrm{NN}}} = 7.7$–200 GeV and Pb+Pb collisions at $\sqrt{s_{\mathrm{NN}}} = 2.76$ and $5.02$ TeV.

At RHIC energies, both scenarios provide a good description of strange hadrons, while the 2CFO scenario achieves improved agreement for $\pi$ and $p$ yields.  At LHC energies, the 2CFO scenario demonstrates a noticeable
overall improvement in agreement with the experimental data across all particle species.

Figure~\ref{fig:Fit_nuclei} presents the results of thermal fits that include light nuclei, performed using both 1CFO and 2CFO in the HRG and HRG-PCE frameworks. Within the HRG scenario, the inclusion of light nuclei reduces the overall agreement of the 1CFO fits, particularly for strange hadrons. In contrast, the 2CFO scenario continues to provide better agreement with the experimental data.

At RHIC energies, in the HRG framework, the 2CFO scenario offers an improved description of $t$ and $\bar{d}$ yields compared to 1CFO, while the $d$ yield remains similarly described in both scenarios. In the HRG-PCE framework, the $d$ yields are better described in both 1CFO and 2CFO relative to HRG. Moreover, $t$ and $\bar{d}$ are best reproduced in the 2CFO scenario within HRG-PCE, emphasizing the importance of not only a flavour-dependent freeze-out but also incorporating post-chemical freeze-out expansion for light nuclei production.

At LHC energies, the 2CFO scenario provides the most accurate and consistent description of hadron and light nuclei yields across both the HRG and HRG-PCE frameworks.

The variations of chemical freeze-out parameters $T_{\text{ch}}$, $\mu^{(0)}_B$, and $R^{(0)}$ with $\sqrt{s_\mathrm{NN}}$ extracted within the 2CFO scenario, with and without light nuclei included in the thermal fits, are shown in Fig. \ref{fig:para_en}. Thermal fit parameters for each energy and particle set in both 1CFO and 2CFO scenarios are listed in Table \ref{table:parameters}.

Including light nuclei in the thermal fit leads to a lower freeze-out temperature and a larger fireball radius in 1CFO scenario, indicating that light nuclei favor a later freeze-out. Consistently, the $T_{\text{ch}}$ of the \textit{light hadrons}(+\textit{nuclei}) set is 10–20 MeV lower than that of the \textit{strange hadrons} set, further supporting a sequential freeze-out scenario where light hadrons and nuclei freeze-out later than strange hadrons. Additionally, we observe that $\mu^{(0)}_B$ is higher for the \textit{strange hadrons} set, with the difference becoming more pronounced at lower energies, consistent with previous findings \cite{Flor:2021olm}. 

Chemical freeze-out parameters are observed to vary only slightly (<5\%) when the thermal fit is performed within the HRG-PCE framework at $T_{\text{kin}} = 100$ MeV compared to the HRG. To facilitate predictions across a continuous range of collision energies, we describe the energy dependence of the chemical freeze-out parameters, $T_{\text{ch}}$, $\mu^{(0)}_B$, and $R^{(0)}$, using smooth functions of $\sqrt{s_\mathrm{NN}}$~\cite{Andronic:2008gu},

\begin{equation}
    \begin{aligned}
        T_{\text{ch}}=T_{\text{ch}}^{\text{lim}} \frac{1}{1+\exp(\frac{w-\ln(\sqrt{s_\mathrm{NN}})}{x})}, \\
        \mu^{(0)}_B = \frac{y}{1+z\sqrt{s_\mathrm{NN}}}, \text{and}\\
        R^{(0)} = q(\sqrt{s_\mathrm{NN}})^r,
        \label{eq:part}
    \end{aligned}
\end{equation}

where $T_{\text{ch}}^{\text{lim}}$ is the limiting temperature. The parameters $T_{\text{ch}}^{\text{lim}}$, $w$, $x$, $y$, $z$, $q$, and $r$  are listed in Table \ref{tab:fit_parameters}.

We use the parametrized values of $T_{\text{ch}}$, $\mu^{(0)}_B$, and $R^{(0)}$ to calculate the light nuclei yield ratios: $d/p$, $\overline{d}/\overline{p}$, $t/p$, $t/d$, $^4\text{He}/^3\text{He}$, and $^3\text{H}_{\Lambda}/^3\text{He}$. These ratios are calculated within both the 1CFO and 2CFO scenarios of the HRG and HRG-PCE frameworks and are contrasted with the corresponding experimental ratios in 0-10\% central Au+Au and Pb+Pb collisions reported by the STAR and ALICE collaborations~\cite{STAR:2019sjh,STAR:2022hbp,STAR:2010gyg,ALICE:2015oer,STAR:2023fbc,ALICE:2024koa}. We note that the $^3\text{H}_{\Lambda}/^3\text{He}$ ratio is a mixed strange-to-nonstrange ratio. In the 2CFO scenario, the $^3\text{H}_{\Lambda}$ yield is calculated using the \textit{strange hadrons} chemical freeze-out parameters, while the $^3$He yield is obtained using the \textit{light hadrons}(+\textit{nuclei}) chemical freeze-out parameters. Figure~\ref{fig:LNRatios} presents this comparison as a function of $\sqrt{s_\mathrm{NN}}$. We observe that, 

\begin{itemize}

\item For the $d/p$ ratio, excluding light nuclei from the thermal fits leads to an overprediction in the 1CFO scenario, while the 2CFO scenario shows better agreement with the data across most energies in both HRG and HRG-PCE frameworks. Including light nuclei in the fit suppresses the ratio, improving consistency with experimental data. The best description is achieved in the HRG-PCE framework with light nuclei included.

\item For the $\overline{d}/\overline{p}$ ratio, omitting light nuclei from the thermal fits results in an overestimation by the 1CFO scenario in both HRG and HRG-PCE, whereas the 2CFO scenario aligns more closely with the data. Agreement improves for both scenarios when light nuclei are included, with the best description obtained in the 2CFO scenario within both HRG and HRG-PCE frameworks.

\item For the $t/p$ ratio, neither the 1CFO nor 2CFO scenario reproduces the data well without including light nuclei. Upon inclusion, the 2CFO scenario, particularly within the HRG-PCE framework, offers the closest agreement with the experimental data within uncertainties.

\item For the $t/d$ ratio, neither scenario captures the data accurately when light nuclei are excluded from the thermal fits. Including them leads to the best agreement in the 2CFO scenario within the HRG-PCE framework, within uncertainties, except for Au+Au collisions at $\sqrt{s_{\mathrm{NN}}} = 11.5$ and 200 GeV.

\item For the $^4\text{He}/^3\text{He}$ ratio, both the 1CFO and 2CFO scenarios within the HRG and HRG-PCE frameworks describe the experimental data within uncertainties, except in Pb+Pb collisions at $\sqrt{s_{\mathrm{NN}}} = 5.02$ TeV, where the 1CFO scenario provides a better description.

\item For the $^3\text{H}_\Lambda/^3\text{He}$ ratio, the 1CFO scenario provides a better description of the data than the 2CFO scenario, within uncertainties. The 1CFO scenario within the HRG-PCE framework provides the best agreement with the data for Pb+Pb collisions at $\sqrt{s_{\mathrm{NN}}} = 5.02$ TeV. 

\end{itemize}

Our findings suggest that the 2CFO scenario with light nuclei included within the HRG-PCE framework provides the most consistent description of light nuclei yields and their ratios. While the 1CFO scenario is occasionally sufficient (notably for the $^3_\Lambda\text{H}/^3\text{He}$ and $^4\text{He}/^3\text{He}$ in Pb+Pb collisions at $\sqrt{s_{\mathrm{NN}}} = 5.02$ TeV), it generally overpredicts light nuclei ratio. This study highlights the importance of a flavour-dependent chemical freeze-out and the inclusion of light nuclei in the thermal fits within the HRG-PCE framework. This becomes particularly relevant when describing light nuclei yields in heavy-ion collisions across a broad range of energies.

\section{Summary}
\label{sec:summary}

Hadron and light nuclei yields in Au+Au collisions at RHIC energies and Pb+Pb collisions at LHC energies were studied using the HRG and HRG-PCE model frameworks, with a focus on flavour-dependent chemical freeze-out. The analysis, performed using the \texttt{Thermal-FIST} package within the Grand Canonical Ensemble (GCE) formalism, incorporated quantum statistics for all particles.

Chemical freeze-out parameters, $T_\text{ch}$, $\mu^{(0)}_B$, and $R^{(0)}$, were extracted under both 1CFO and 2CFO scenarios by fitting yields of various hadrons and light nuclei. We systematically examined the impact of including light nuclei in thermal fits, as well as the effects of post–chemical freeze-out evolution within the HRG-PCE framework, where light nuclei yields are computed at $T_{\text{kin}} = 100$ MeV based on hadron yields fixed at $T_{\text{ch}}$.

The resulting chemical freeze-out parameters suggest a flavour-dependent chemical freeze-out, where strange hadrons freeze-out earlier than light hadrons at all collision energies. Additionally, including light nuclei in the thermal fits leads to lower extracted freeze-out temperatures and larger fireball radii, suggesting that light nuclei favour a later freeze-out.

To further understand the role of flavour-dependent chemical freeze-out on light nuclei yields, we compare the calculated yield ratios, $d/p$, $\bar{d}/\bar{p}$, $t/p$, $t/d$, $^4\text{He}/^3\text{He}$, and $^3\text{H}_{\Lambda}/^3\text{He}$, from both 1CFO and 2CFO scenarios to experimental measurements. The 2CFO scenario consistently provides a better description of these ratios, with the HRG-PCE framework offering the best agreement in most cases, except for the $^3\text{H}_{\Lambda}/^3\text{He}$ and $^4\text{He}/^3\text{He}$ in Pb+Pb collisions at $\sqrt{s_{\mathrm{NN}}} = 5.02$ TeV, where 1CFO performs better.

Overall, our findings support a flavour-dependent chemical freeze-out scenario at RHIC and LHC energies, suggesting that strange and light hadrons, including light nuclei, may freeze-out at different stages of the hadronic evolution. This study highlights the importance of accounting for flavour-dependent chemical freeze-out and post-chemical freeze-out effects while interpreting final-state light nuclei yields using thermal models in heavy-ion collisions.

\section*{Acknowledgments}
CJ acknowledges the financial support from DAE-DST, Government of India bearing Project No. SR/MF/PS-02/2021-IISERT (E-37130). FAF and HC acknowledge the DOE grant DE-SC004168 for supporting this work. Additionally, FAF would like to thank the National Science Foundation Grant No. 2138010.

 \bibliographystyle{spphys}
 \bibliography{refrences.bib}

\begin{thebibliography}{10}
\providecommand{\url}[1]{{#1}}
\providecommand{\urlprefix}{URL }
\expandafter\ifx\csname urlstyle\endcsname\relax
  \providecommand{\doi}[1]{DOI \discretionary{}{}{}#1}\else
  \providecommand{\doi}{DOI \discretionary{}{}{}\begingroup \urlstyle{rm}\Url}\fi

\bibitem{STAR:2005gfr}
J.~Adams, et~al., Nucl. Phys. A \textbf{757}, 102 (2005).
\newblock \doi{10.1016/j.nuclphysa.2005.03.085}

\bibitem{Aoki:2006we}
Y.~Aoki, G.~Endrodi, Z.~Fodor, S.D. Katz, K.K. Szabo, Nature \textbf{443}, 675 (2006).
\newblock \doi{10.1038/nature05120}

\bibitem{Ejiri:2008xt}
S.~Ejiri, Phys. Rev. D \textbf{78}, 074507 (2008).
\newblock \doi{10.1103/PhysRevD.78.074507}

\bibitem{ALICE:2022wpn}
S.~Acharya, et~al., Eur. Phys. J. C \textbf{84}(8), 813 (2024).
\newblock \doi{10.1140/epjc/s10052-024-12935-y}

\bibitem{Chatterjee:2015fua}
S.~Chatterjee, S.~Das, L.~Kumar, D.~Mishra, B.~Mohanty, R.~Sahoo, N.~Sharma, Adv. High Energy Phys. \textbf{2015}, 349013 (2015).
\newblock \doi{10.1155/2015/349013}

\bibitem{Cleymans:1998fq}
J.~Cleymans, K.~Redlich, Phys. Rev. Lett. \textbf{81}, 5284 (1998).
\newblock \doi{10.1103/PhysRevLett.81.5284}

\bibitem{Becattini:2009sc}
F.~Becattini, in \emph{{International School on Quark-Gluon Plasma and Heavy Ion Collisions: past, present, future}} (2009)

\bibitem{Andronic:2017pug}
A.~Andronic, P.~Braun-Munzinger, K.~Redlich, J.~Stachel, Nature \textbf{561}(7723), 321 (2018).
\newblock \doi{10.1038/s41586-018-0491-6}

\bibitem{Chatterjee:2013yga}
S.~Chatterjee, R.M. Godbole, S.~Gupta, Phys. Lett. B \textbf{727}, 554 (2013).
\newblock \doi{10.1016/j.physletb.2013.11.008}

\bibitem{Bugaev:2013sfa}
K.A. Bugaev, D.R. Oliinychenko, J.~Cleymans, A.I. Ivanytskyi, I.N. Mishustin, E.G. Nikonov, V.V. Sagun, EPL \textbf{104}(2), 22002 (2013).
\newblock \doi{10.1209/0295-5075/104/22002}

\bibitem{Chatterjee:2014ysa}
S.~Chatterjee, B.~Mohanty, Phys. Rev. C \textbf{90}(3), 034908 (2014).
\newblock \doi{10.1103/PhysRevC.90.034908}

\bibitem{Ratti:2011au}
C.~Ratti, R.~Bellwied, M.~Cristoforetti, M.~Barbaro, Phys. Rev. D \textbf{85}, 014004 (2012).
\newblock \doi{10.1103/PhysRevD.85.014004}

\bibitem{Bellwied:2013cta}
R.~Bellwied, S.~Borsanyi, Z.~Fodor, S.D. Katz, C.~Ratti, Phys. Rev. Lett. \textbf{111}, 202302 (2013).
\newblock \doi{10.1103/PhysRevLett.111.202302}

\bibitem{Flor:2020fdw}
F.A. Flor, G.~Olinger, R.~Bellwied, Phys. Lett. B \textbf{814}, 136098 (2021).
\newblock \doi{10.1016/j.physletb.2021.136098}

\bibitem{Flor:2021olm}
F.A. Flor, G.~Olinger, R.~Bellwied, Phys. Lett. B \textbf{834}, 137473 (2022).
\newblock \doi{10.1016/j.physletb.2022.137473}

\bibitem{Andronic:2010qu}
A.~Andronic, P.~Braun-Munzinger, J.~Stachel, H.~Stocker, Phys. Lett. B \textbf{697}, 203 (2011).
\newblock \doi{10.1016/j.physletb.2011.01.053}

\bibitem{Yu:2024sof}
N.~Yu, Z.~Zhang, H.~Xu, M.X. Song, M.~Song, Nucl. Sci. Tech. \textbf{36}(4), 65 (2025).
\newblock \doi{10.1007/s41365-025-01661-z}

\bibitem{Oliinychenko:2020ply}
D.~Oliinychenko, Nucl. Phys. A \textbf{1005}, 121754 (2021).
\newblock \doi{10.1016/j.nuclphysa.2020.121754}

\bibitem{Siemens:1979dz}
P.J. Siemens, J.I. Kapusta, Phys. Rev. Lett. \textbf{43}, 1486 (1979).
\newblock \doi{10.1103/PhysRevLett.43.1486}

\bibitem{Hahn:1986mb}
D.~Hahn, H.~Stoecker, Nucl. Phys. A \textbf{476}, 718 (1988).
\newblock \doi{10.1016/0375-9474(88)90332-6}

\bibitem{STAR:2019sjh}
J.~Adam, et~al., Phys. Rev. C \textbf{99}(6), 064905 (2019).
\newblock \doi{10.1103/PhysRevC.99.064905}

\bibitem{STAR:2022hbp}
M.~Abdulhamid, et~al., Phys. Rev. Lett. \textbf{130}, 202301 (2023).
\newblock \doi{10.1103/PhysRevLett.130.202301}

\bibitem{Sun:2022xjr}
K.J. Sun, R.~Wang, C.M. Ko, Y.G. Ma, C.~Shen, Nature Commun. \textbf{15}(1), 1074 (2024).
\newblock \doi{10.1038/s41467-024-45474-x}

\bibitem{Motornenko:2019jha}
A.~Motornenko, V.~Vovchenko, C.~Greiner, H.~Stoecker, Phys. Rev. C \textbf{102}(2), 024909 (2020).
\newblock \doi{10.1103/PhysRevC.102.024909}

\bibitem{Vovchenko:2019aoz}
V.~Vovchenko, K.~Gallmeister, J.~Schaffner-Bielich, C.~Greiner, Phys. Lett. B \textbf{800}, 135131 (2020).
\newblock \doi{10.1016/j.physletb.2019.135131}

\bibitem{Vovchenko:2018fmh}
V.~Vovchenko, M.I. Gorenstein, H.~Stoecker, Phys. Rev. C \textbf{98}(3), 034906 (2018).
\newblock \doi{10.1103/PhysRevC.98.034906}

\bibitem{Vovchenko:2019pjl}
V.~Vovchenko, H.~Stoecker, Comput. Phys. Commun. \textbf{244}, 295 (2019).
\newblock \doi{10.1016/j.cpc.2019.06.024}

\bibitem{Garcilazo:1987rx}
H.~Garcilazo, Phys. Rev. C \textbf{35}, 1820 (1987).
\newblock \doi{10.1103/PhysRevC.35.1820}

\bibitem{ParticleDataGroup:2020ssz}
P.A. Zyla, et~al., PTEP \textbf{2020}(8), 083C01 (2020).
\newblock \doi{10.1093/ptep/ptaa104}

\bibitem{STAR:2004yym}
J.~Adams, et~al., Phys. Lett. B \textbf{612}, 181 (2005).
\newblock \doi{10.1016/j.physletb.2004.12.082}

\bibitem{STAR:2019bjj}
J.~Adam, et~al., Phys. Rev. C \textbf{102}(3), 034909 (2020).
\newblock \doi{10.1103/PhysRevC.102.034909}

\bibitem{STAR:2017sal}
L.~Adamczyk, et~al., Phys. Rev. C \textbf{96}(4), 044904 (2017).
\newblock \doi{10.1103/PhysRevC.96.044904}

\bibitem{Bellini:2018khg}
F.~Bellini, Nucl. Phys. A \textbf{982}, 427 (2019).
\newblock \doi{10.1016/j.nuclphysa.2018.09.082}

\bibitem{ALICE:2022veq}
S.~Acharya, et~al., Phys. Rev. C \textbf{107}(6), 064904 (2023).
\newblock \doi{10.1103/PhysRevC.107.064904}

\bibitem{ALICE:2019hno}
S.~Acharya, et~al., Phys. Rev. C \textbf{101}(4), 044907 (2020).
\newblock \doi{10.1103/PhysRevC.101.044907}

\bibitem{ALICE:2014jbq}
B.B. Abelev, et~al., Phys. Rev. C \textbf{91}, 024609 (2015).
\newblock \doi{10.1103/PhysRevC.91.024609}

\bibitem{ALICE:2023ulv}
S.~Acharya, et~al., Phys. Rev. Lett. \textbf{133}(9), 092301 (2024).
\newblock \doi{10.1103/PhysRevLett.133.092301}

\bibitem{STAR:2008med}
B.I. Abelev, et~al., Phys. Rev. C \textbf{79}, 034909 (2009).
\newblock \doi{10.1103/PhysRevC.79.034909}

\bibitem{STAR:2011fbd}
G.~Agakishiev, et~al., Phys. Rev. Lett. \textbf{108}, 072301 (2012).
\newblock \doi{10.1103/PhysRevLett.108.072301}

\bibitem{Magestro:2001jz}
D.~Magestro, J. Phys. G \textbf{28}, 1745 (2002).
\newblock \doi{10.1088/0954-3899/28/7/328}

\bibitem{Schnedermann:1993ws}
E.~Schnedermann, J.~Sollfrank, U.W. Heinz, Phys. Rev. C \textbf{48}, 2462 (1993).
\newblock \doi{10.1103/PhysRevC.48.2462}

\bibitem{ALICE:2015wav}
J.~Adam, et~al., Phys. Rev. C \textbf{93}(2), 024917 (2016).
\newblock \doi{10.1103/PhysRevC.93.024917}

\bibitem{NA49:2016qvu}
T.~Anticic, et~al., Phys. Rev. C \textbf{94}(4), 044906 (2016).
\newblock \doi{10.1103/PhysRevC.94.044906}

\bibitem{Andronic:2008gu}
A.~Andronic, P.~Braun-Munzinger, J.~Stachel, Phys. Lett. B \textbf{673}, 142 (2009).
\newblock \doi{10.1016/j.physletb.2009.06.021}.
\newblock [Erratum: Phys.Lett.B 678, 516 (2009)]

\bibitem{STAR:2010gyg}
B.I. Abelev, et~al., Science \textbf{328}, 58 (2010).
\newblock \doi{10.1126/science.1183980}

\bibitem{ALICE:2015oer}
J.~Adam, et~al., Phys. Lett. B \textbf{754}, 360 (2016).
\newblock \doi{10.1016/j.physletb.2016.01.040}

\bibitem{STAR:2023fbc}
M.~Abdulhamid, et~al., Nature \textbf{632}(8027), 1026 (2024).
\newblock \doi{10.1038/s41586-024-07823-0}

\bibitem{ALICE:2024koa}
S.~Acharya, et~al., Phys. Lett. B \textbf{860}, 139066 (2025).
\newblock \doi{10.1016/j.physletb.2024.139066}

\end{thebibliography}

 \appendix
\setcounter{table}{0}
\renewcommand{\thetable}{A\arabic{table}}
\begin{table*}
\centering
\caption{Chemical freeze-out parameters for different freeze-out scenarios at various collision energies.}
\setlength{\tabcolsep}{20pt} 
\renewcommand{\arraystretch}{1.1} 
\begin{tabular}{ccccc}
\hline
$\sqrt{s_\mathrm{NN}}$ (GeV) & $T_\text{ch}$ (MeV) & $\mu^{(0)}_B$ (MeV) & $R^{(0)}$ (fm) & $\chi^2/\mathrm{ndf}$ \\
\hline
\multicolumn{5}{c}{1CFO \textit{hadrons}: $\pi$, $K$, $K_s^{0}$, $p$, $\phi$, $\Lambda$, $\Xi$, and $\Omega$} \\
\hline
7.7   & $149.55 \pm 1.70$ & $435.88 \pm 11.97$ & $5.10 \pm 0.18$ & $22.5/9$ \\
11.5  & $158.66 \pm 1.58$ & $330.13 \pm 10.52$ & $5.00 \pm 0.15$ & $34.3/11$ \\
19.6  & $165.16 \pm 1.34$ & $214.65 \pm 5.99$  & $5.14 \pm 0.13$ & $25.6/11$ \\
27    & $166.66 \pm 1.30$ & $161.71 \pm 5.86$  & $5.26 \pm 0.12$ & $31.7/11$ \\
39    & $165.83 \pm 1.84$ & $112.40 \pm 8.34$  & $5.56 \pm 0.18$ & $16.0/11$ \\
200   & $162.97 \pm 1.85$ & $20.45 \pm 10.26$  & $7.20 \pm 0.23$ & $27.8/10$ \\
2760  & $156.69 \pm 1.99$ & $1.00$    & $10.19 \pm 0.38$ & $17.7/12$ \\
5020  & $156.14 \pm 1.63$ & $1.00$    & $10.70 \pm 0.31$ & $70.0/12$ \\
\hline
\multicolumn{5}{c}{1CFO \textit{hadrons}+\textit{nuclei}: $\pi$, $K$, $K_s^{0}$, $p$, $\phi$, $\Lambda$, $\Xi$, $\Omega$, $d$, and $t$ or $^3\text{He}$} \\
\hline
7.7   & $141.58 \pm 1.01$ & $371.95 \pm 3.36$  & $6.03 \pm 0.12$ & $76.8/11$ \\
11.5  & $150.28 \pm 0.86$ & $269.21 \pm 2.97$  & $5.88 \pm 0.10$ & $99.7/14$ \\
19.6  & $155.60 \pm 0.73$ & $174.15 \pm 2.37$  & $6.11 \pm 0.09$ & $138.0/14$ \\
27    & $157.32 \pm 0.68$ & $129.71 \pm 2.29$  & $6.15 \pm 0.08$ & $139.4/14$ \\
39    & $155.48 \pm 0.85$ & $95.12 \pm 2.75$   & $6.61 \pm 0.12$ & $75.6/14$ \\
200   & $151.31 \pm 0.77$ & $16.99 \pm 2.73$   & $8.69 \pm 0.15$ & $94.6/13$ \\
2760  & $153.74 \pm 1.29$ & $1.00$    & $10.73 \pm 0.28$ & $23.2/16$ \\
5020  & $149.93 \pm 0.41$ & $1.00$    & $11.95 \pm 0.13$ & $106.6/16$ \\
\hline
\multicolumn{5}{c}{2CFO \textit{light hadrons}: $\pi$, $K$, and $p$} \\
\hline	
7.7   & $138.75 \pm 7.94$ & $404.80 \pm 14.79$ & $6.34 \pm 0.85$ & $1.5/3$ \\
11.5  & $146.45 \pm 8.60$ & $300.55 \pm 18.29$ & $6.36 \pm 0.87$ & $2.8/3$ \\
19.6  & $149.22 \pm 9.08$ & $195.66 \pm 21.24$ & $6.80 \pm 0.95$ & $4.4/3$ \\
27    & $150.13 \pm 8.48$ & $152.88 \pm 20.85$ & $6.96 \pm 0.91$ & $3.2/3$ \\
39    & $151.32 \pm 8.18$ & $106.59 \pm 20.68$ & $7.01 \pm 0.87$ & $3.1/3$ \\
200   & $160.64 \pm 5.41$ & $25.31 \pm 14.26$  & $7.50 \pm 0.61$ & $3.9/3$ \\
2760  & $149.26 \pm 3.15$ & $1.00$    & $11.47 \pm 0.63$ & $4.5/4$ \\
5020  & $147.94 \pm 1.90$ & $1.00$    & $12.19 \pm 0.40$ & $11.4/4$ \\
\hline
\multicolumn{5}{c}{2CFO \textit{light hadrons}+\textit{nuclei}: $\pi$, $p$, $d$, and $t$ or $^3\text{He}$} \\	
\hline
7.7   & $127.69 \pm 6.69$ & $394.69 \pm 18.73$ & $7.75 \pm 0.77$ & $5.9/3$ \\
11.5  & $140.93 \pm 3.73$ & $280.60 \pm 11.67$ & $7.19 \pm 0.41$ & $9.8/4$ \\
19.6  & $147.13 \pm 1.70$ & $183.02 \pm 5.59$  & $7.31 \pm 0.26$ & $6.0/4$ \\
27    & $149.31 \pm 1.32$ & $138.08 \pm 4.15$  & $7.34 \pm 0.23$ & $7.8/4$ \\
39    & $150.89 \pm 1.20$ & $98.06 \pm 3.56$   & $7.32 \pm 0.23$ & $9.4/4$ \\
200   & $148.82 \pm 0.93$ & $19.57 \pm 3.23$   & $9.11 \pm 0.21$ & $16.7/4$ \\
2760  & $149.68 \pm 1.84$ & $1.00$    & $11.57 \pm 0.44$ & $2.0/6$ \\
5020  & $148.70 \pm 0.51$ & $1.00$    & $12.40 \pm 0.19$ & $21.2/6$ \\
\hline
\multicolumn{5}{c}{2CFO \textit{strange hadrons}: $K$, $K_s^{0}$, $\phi$, $\Lambda$, $\Xi$, and $\Omega$} \\
\hline
7.7   & $150.49 \pm 1.77$ & $442.57 \pm 13.56$ & $4.98 \pm 0.18$ & $14.6/5$ \\
11.5  & $159.95 \pm 1.65$ & $337.45 \pm 11.61$ & $4.86 \pm 0.16$ & $19.8/7$ \\
19.6  & $166.21 \pm 1.39$ & $218.23 \pm 6.32$  & $5.03 \pm 0.13$ & $10.5/7$ \\
27    & $167.53 \pm 1.34$ & $163.60 \pm 6.15$  & $5.16 \pm 0.12$ & $18.4/7$ \\
39    & $167.67 \pm 1.99$ & $116.81 \pm 9.47$  & $5.36 \pm 0.19$ & $6.5/7$ \\
200   & $164.96 \pm 2.15$ & $14.55 \pm 14.83$  & $6.90 \pm 0.27$ & $22.7/6$ \\
2760  & $161.06 \pm 2.52$ & $1.00$    & $9.40 \pm 0.45$ & $5.6/8$ \\
5020  & $167.89 \pm 2.23$ & $1.00$    & $8.94 \pm 0.32$ & $11.2/8$ \\
\hline
\end{tabular}
\label{table:parameters}
\end{table*}

\end{document}